\documentclass[twocolumn,a4paper]{aa}

\usepackage[suffix=]{epstopdf}
\begin{document}

\title{Hard X-ray emitting energetic electrons and photospheric electric currents}
\author{S. Musset
\and N. Vilmer
\and V. Bommier}
\institute{LESIA, Observatoire de Paris, CNRS, UPMC, Universit\'{e} Paris-Diderot, 5 place Jules Janssen, 92195 Meudon, France}
\date{Received ... / Accepted ...}

\abstract {The energy released during solar flares is believed to be stored in non-potential magnetic fields associated with electric currents flowing in the corona. While no measurements of coronal electric currents are presently available, maps of photospheric electric currents can now be derived from SDO/HMI observations. Photospheric electric currents have been shown to be the tracers of the coronal electric currents. Particle acceleration can result from electric fields associated with coronal electric currents. We revisit here some aspects of the relationship between particle acceleration in solar flares and electric currents in the active region.}
{We study the relation between the energetic electron interaction sites in the solar atmosphere, and the magnitudes and changes of vertical electric current densities measured at the photospheric level, during the X2.2 flare on February 15 2011 in AR NOAA 11158.}
{X-ray images from the \textit{Reuven Ramaty High Energy Solar Spectroscopic Imager} (RHESSI) are overlaid on magnetic field and electric current density maps calculated from the spectropolarimetric measurements of the \textit{Helioseismic and Magnetic Imager} (HMI) on the \textit{Solar Dynamics Observatory} (SDO) using the UNNOFIT inversion and Metcalf disambiguation codes. X-ray images are also compared with Extreme Ultraviolet (EUV) images from the SDO \textit{Atmospheric Imaging Assembly} (AIA) to complement the flare analysis.}
{Part of the elongated X-ray emissions from both thermal and non-thermal electrons overlay the elongated narrow current ribbons observed at the photospheric level. A new X-ray source at 50-100 keV (produced by non-thermal electrons) is observed in the course of the flare and is cospatial with a region in which new vertical photospheric currents appeared during the same period (increase of 15\%). These observational results are discussed in the context of the scenarios in which magnetic reconnection (and subsequent plasma heating and particle acceleration) occurs at current-carrying layers in the corona.}
{}

\keywords{Sun: flares - Sun: particle emission - Sun: X-rays - Sun: magnetic fields - Acceleration of particles - RHESSI - SDO/AIA - SDO/HMI}

\maketitle

\section{Introduction}

It is now commonly admitted that solar flares are the result of the sudden release of magnetic free energy in the corona, this free energy being stored in non-potential magnetic fields associated with electric currents flowing in the corona.
As there are currently no measurements of the vector magnetic fields in the corona, there are also no measurements of coronal electric currents. Electric currents have been only determined in a few cases at the photospheric level using vector magnetic field measurements achieved in photospheric lines.
While some pioneering work was done in the 1970's by e.g. \cite{moreton_severny} at the Crimean Astrophysical Observatory, many observations of vector magnetic fields and related electric currents have been obtained later in the 1980's and 1990's with ground-based vector magnetographs \citep[see e.g.][]{hagyard_et_al_1984,canfield_et_al_1993}.
Nowadays, the polarimetric measurements obtained continuously, and with an unprecedented cadence and spatial resolution with the \textit{Helioseismic and Magnetic Imager} (HMI) on \textit{Solar Dynamics Observatory} (SDO) allows to obtain maps of photospheric vertical electric currents for flaring active regions and even more to study the evolution of these currents on a time resolution of 12 minutes \citep[see e.g.][]{petrie_2012,petrie_2013,janvier}.

The relation between the energetic phenomena (plasma heating and particle acceleration) occurring in solar flares and the electric current systems in active regions has been the subject of many studies for several decades.
In particular, the link between the photospheric electric currents and the energetic electron precipitation sites has been investigated in several events, originally to discuss the relevance of the different flare models presented in the literature.
The first studies used H$\alpha$ observations to characterize electron precipitation sites since no HXR imaging observations were available.
Using data from the Crimean Astrophysical Observatory, \cite{moreton_severny} examined the spatial relationship between H$\alpha$ flare kernels and maxima of vertical electric current densities.
For 80\% of the 25 events studied, a spatial coincidence (< 6'') was found between the location of the center of the bright H$\alpha$ kernel and the location of the strongest electric current densities (> 8 mA/m$^{2}$).
Such a spatial coincidence between H$\alpha$ flare kernels and strong electric current densities was later confirmed by other studies.
\cite{lin_gaizauskas} found in particular that the sites of strongest H$\alpha$ emissions were cospatial with the vertical current systems at the photospheric level, to within 2''.
However, it was also found that even if some of the flare kernels were close to the electric currents, many of them appeared near the edges of currents \citep{romanov_tsap}.
In a series of papers, \cite{canfield_et_al_1993,leka_et_al_1993,beaujardiere_et_al_1993} revisited the question of the relation between electric current systems and electron precipitation sites using vector magnetograph data from the Mees Solar Observatory and information from H$\alpha$ line profiles which allow to disentangle H$\alpha$ signal from electron precipitation and from high pressure.
Studies of several flares \citep{leka_et_al_1993,beaujardiere_et_al_1993} confirmed that sites of intense nonthermal electron precipitation do not coincide with the regions of strongest vertical currents at the photosphere, but tend to occur on the shoulders of channels of high vertical current density rather than at the vertical current density maxima.
In a further study, \cite{demoulin_1997} found for several events that flare ribbons were located in the vicinity of strong electric currents in the photosphere.
They furthermore derived the locations of the photospheric footprints  of regions of rapid change in magnetic line connectivity called quasi-separatrix layers (QSLs).
They found as in other observational studies \citep[see e.g][]{machado_et_al_1983,mandrini_et_al_1995,bagala_et_al_1995} that flares tend to occur near the locations of these QSLs.
These QSLs are now known to be regions where strong current densities can develop and where reconnection can occur \citep[see e.g.][]{demoulin_1996,aulanier_et_al_2005} thus establishing a clear link between flare energy release, electric currents and magnetic reconnection.

With the arrival of HXR imagers, the link between vertical electric currents at the photosphere and electron precipitation sites has been revisited using in particular YOHKOH/HXT observations \citep[see][]{canfield_et_al_1992,li}.
The latter study based on six events confirms that HXR emission and thus electron precipitation sites are not exactly co-spatial with regions of highest vertical current densities at photospheric levels but are rather adjacent to the current channels, therefore confirming the results obtained by \cite{beaujardiere_et_al_1993} with H$\alpha$ observations.
In a later paper, \cite{aschwanden_et_al_1999} showed how this offset between the maximum of vertical currents and the HXR loop footpoints could be explained in the context of the 3D geometry of quadrupolar reconnection.
It must finally be noticed that in the earlier studies, vector magnetograms (and thus electric current maps) were sparsely derived and that the integration time to derive such information could be more than one hour.
Several hours may also separate the time of the flare and the time of the magnetic field measurements.
Nowadays, the combination of polarimetric measurements continuously obtained with the \textit{Helioseismic and Magnetic Imager} (HMI) on \textit{Solar Dynamics Observatory} (SDO) and of HXR observations of solar flares obtained with the \textit{Reuven Ramaty High Energy Solar Spectroscopic Imager} (RHESSI) allows to compare high quality vector magnetic field (and electric current) maps at a cadence of 12 minutes and HXR observations of solar flares at exactly the same time. This provides new possibilities to study more systematically and in more details the relation between the location of HXR sources (and electron precipitation sites) and the vertical electric current densities.

This paper presents the first results of such a study for the flare of February 15, 2011. The simultaneous evolutions of electric currents and of HXR emission sources during the flare are furthermore examined for the first time.
The February 15, 2011 event is quite appropriate for such a first study. Indeed, this major flare (GOES class X2.2) occurred in the active region AR11158 when the active region was near disk-center.
The derivation of vector magnetic field maps and of vertical electric current densities requires indeed observations of an active region located near the center of the solar disk to have good measurements of the vector magnetic field at the photospheric level and to avoid projection effects.
The February 15, 2011 flare which is the first X-class flare of cycle 24 is the subject of numerous studies since the original analysis by \cite{schrijver_2011} \citep[see e.g. for some of the most recent papers][for MHD simulations of the flare]{inoue_2014}.
In close relationship with the present paper, \cite{janvier} quantified the vertical currents at the photospheric level and studied their temporal evolution during the flare. They concluded in particular that the evolution of both current and flare ribbons is related to the evolution of the magnetic field during the flare in the context of a 3D standard flare model. The active region itself and its activity  has been also extensively studied. In particular, \cite{sun,vemareddy,aschwanden,zhao_2014} realized different estimations of the e.g. free magnetic energy and dissipated energy for several events occurring in this active region.

In section \ref{section2} of this paper, we present briefly the different instruments used in this study as well as the techniques necessary to obtain maps of vector magnetic fields and electric currents from the measurements. In section \ref{section3}, X-ray and EUV observations of the flare are  compared with the vertical electric current densities. Section \ref{section4} discusses the observational results and presents our interpretation. A summary of the paper and of the conclusions is finally presented in section \ref{section5}.

%======================================================================== DATA AND METHODOLOGY ===============================================================================================

\section{ Magnetic field measurements and Flare observations}
\label{section2}

\subsection{Magnetic Field Measurements and Electric Current Calculations}

\begin{figure}
\begin{center}
\includegraphics[width=\linewidth]{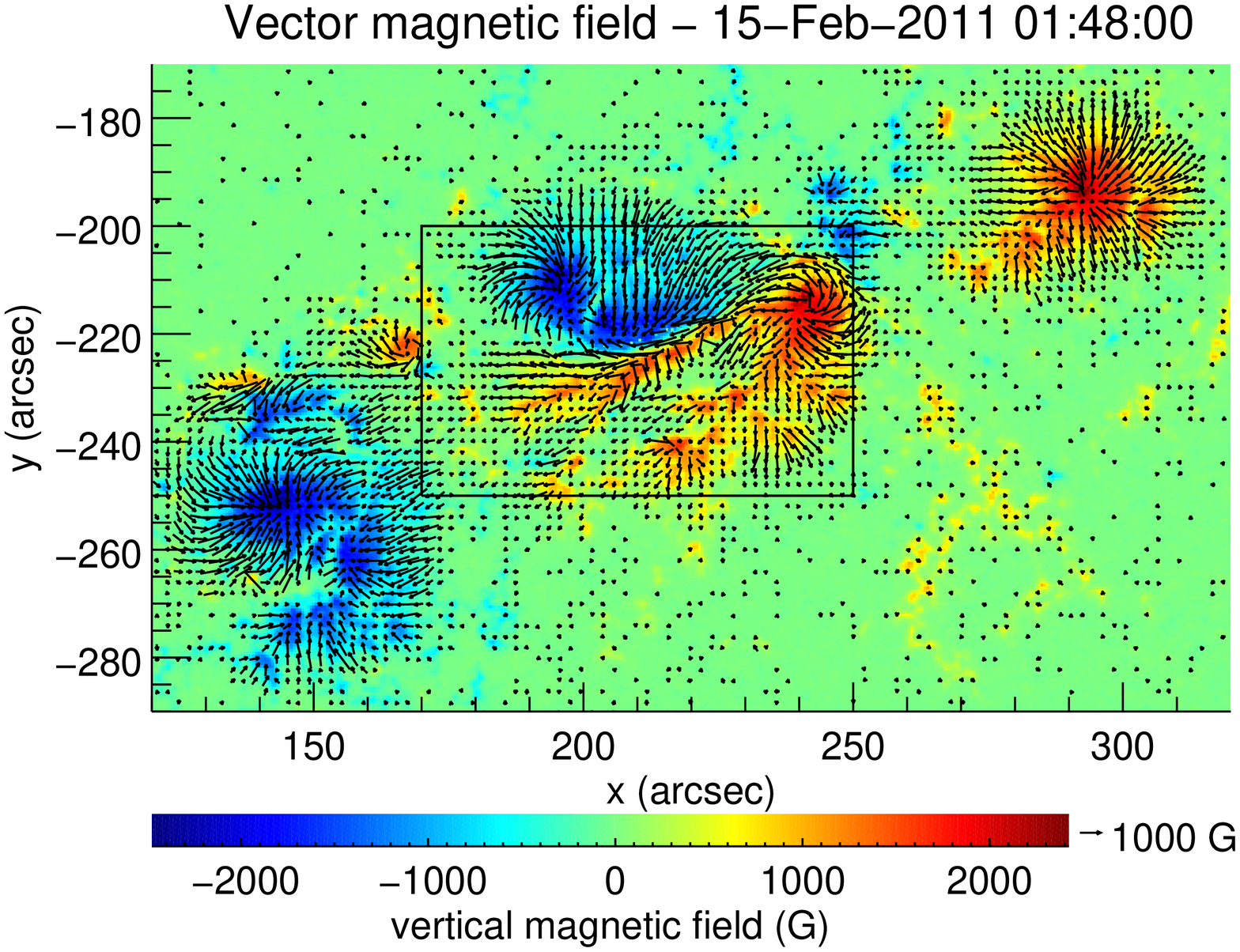}\\
\includegraphics[width=\linewidth]{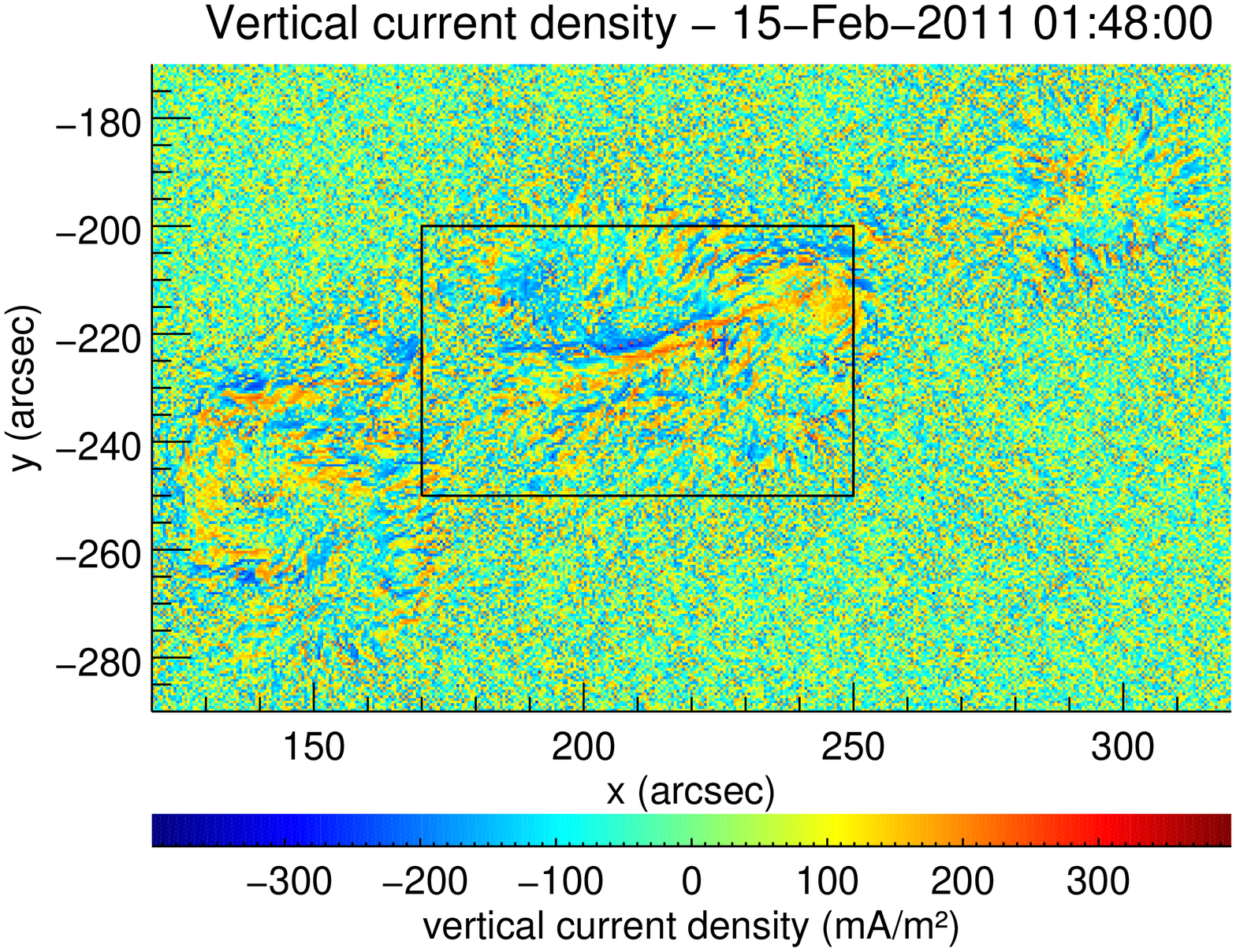}
\end{center}
\caption{Top: vector magnetic field map, bottom: vertical electric current density map of the active region NOAA AR 11158; produced from HMI data, on February 15 2011 at 01:48:00 UT, and represented in the plane-of-sky (see text and Appendix \ref{changeofframe} for more details). The rectangular box represents the field-of-view used in the present study. Top: the arrows represent the horizontal component of the field (for horizontal component greater than 100 G), and the colors represent the intensity of the vertical magnetic field (see scale). Bottom: vertical component of the electric current density (see color scale).}
\label{magn_map}
\end{figure}

The magnetic field and electric current density maps are derivated from the data of the \textit{Helioseismic and Magnetic Imager} HMI on SDO \citep{scherrer,schou}. The HMI instrument provides images of the entire Sun in six narrow spectral bands in a single iron line (FeI 617.33 nm) and in four different states of polarization. This set of 24 images provides spectropolarimetric data to enable the calculation of the full vector magnetic field (inverse problem), at the altitude of formation of the spectral line, which is at the photospheric level in this case.
The three components of the photospheric vector magnetic field can be calculated together with the vertical component of the electric current density (see e.g. figure \ref{magn_map} bottom and \citealt{janvier} for further details). However, because the spectropolarimetric measurements are done in only one single line (i.e. at one altitude), it is not possible to derive all the components of the electric current density but only the vertical one.
%Because the spectropolarimetric measurements are done in only one single line, the full vector magnetic field (three components) can be calculated, but only the vertical component of the current density is accessible.
The spatial resolution is 0.91 arcsec %\citep{scherrer,schou}
and one map of the vector magnetic field and of the vertical electric current can be calculated at a 12 minute-time cadence.

The details of the calculation of magnetic field and vertical density for this set of data is described in \cite{janvier}.
Level-1b IQUV data were used for inversion with the Milne-Eddington inversion code UNNOFIT presented in \cite{Bommier-etal-07}.
The specificity of UNNOFIT is that, in order to take into account the unresolved magnetic structures, a magnetic filling factor is introduced as a free parameter of the Levenberg-Marquardt algorithm that fits the observed set of profiles with a theoretical one. However, for further application we use only the averaged field, i.e. the product of the field by the magnetic filling factor, as recommended by \citet{Bommier-etal-07}.
The interest of the method lies in a better determination of the field inclination. This is of major importance for our study since the currents are determined from the horizontal components of the magnetic field.

After the inversion, the $180^{\circ}$ remaining azimuth ambiguity was resolved by applying the ME0 code developed by Metcalf, Leka, Barnes and Crouch \citep{Leka-etal-09} and available at http://www.cora.nwra.com/AMBIG/. After resolving the ambiguity, the magnetic field vectors were rotated into the local reference frame (i.e. in the heliographic reference frame, see figure \ref{geometry}), where the local vertical axis is the $Oz$ axis (perpendicular to the solar surface at the center of the frame).
The vertical component of the electric current density (perpendicular to the plane of photosphere, i.e. in the heliographic coordinates) was then calculated via the curl of the magnetic field.

\subsection{Flare observations}

RHESSI (\textit{Reuven Ramaty High Energy Solar Spectroscopic Imager}) measures hard X-ray and gamma-ray emissions from the Sun in the 3 keV - 10 MeV range \citep{rhessi}.
Equipped with 9 rotating collimators (pairs of identical grids in front of each detector), it provides images \citep{rhessi_imaging} in addition to spectra \citep{rhessi_spectro}.
The spatial resolution is determined by the pitch of the grids of the collimators used to reconstruct the image. In this paper, images were obtained using the CLEAN and Visibility Forward Fit algorithms using all the collimators except the first one (smallest pitch): the spatial resolution is then of 3.9 arcsec.

The AIA \textit{Atmospheric Imaging Assembly} instrument \citep{aia} aboard SDO (\textit{Solar Dynamics Observatory}) provides images of the solar chromosphere and corona in EUV in six different wavelengths.
The spatial resolution is 1.5 arcsec and for each wavelength, the time cadence is of 12 seconds, which enables the detailed study of the evolution in the flare of magnetic structures and flaring plasma.
%AIA images provide informations about coronal structures where the plasma is denser and hotter than the plasma of the ambiant medium and the high resolution enables the detailed study of the evolution of such magnetic structures.
As the X2.2 flare on February 15 2011 is very intense, most of the AIA channels are saturated around the time of the peak of the flare, even with the minimal exposure time.
Only the channels at 94 \AA \ and 335 \AA \ are not saturated at the time of the flare and therefore used for comparison with X-ray and magnetic observations.

While magnetic fields and current densities are calculated in the heliographic coordinates, RHESSI and AIA images are obtained in the plane-of-sky.
Therefore, to combine e.g. magnetic field maps with e.g. RHESSI images, it is necessary to represent the magnetic fields and current densities in the plane-of-sky coordinates. Figure \ref{geometry} shows the geometry associated with the different measurements and the change of coordinates to compare the different observations is described in Appendix \ref{changeofframe}.
One map of the full vector magnetic field and of the vertical currents represented in the plane-of-sky is shown in figure \ref{magn_map} (note that the components of the field and vertical current density are not changed in this process).

%======================================================================= 3 = OBSERVATIONS - DESCRIPTION ===============================================================================================

\section{Observational Results}
\label{section3}

\subsection{Temporal and spatial evolution of the X-ray emissions}
\label{obs_hxr}

\begin{figure}
\begin{center}
\includegraphics[width=\linewidth]{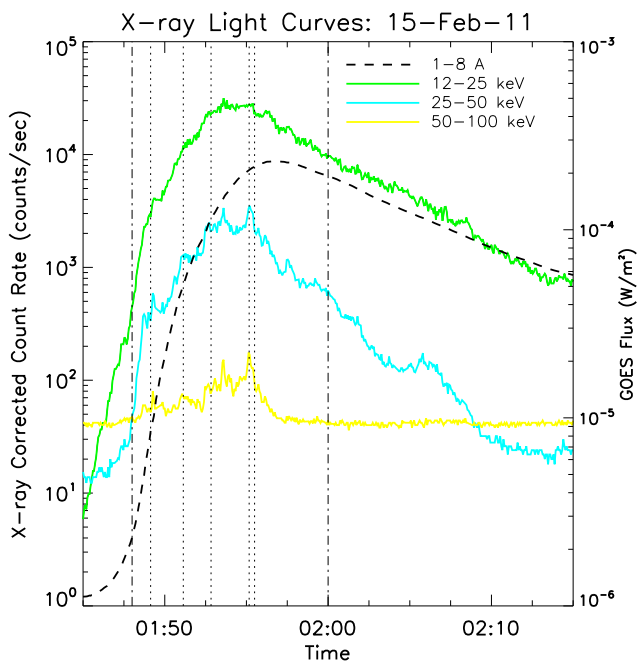}
\end{center}
\caption{RHESSI corrected count rates between 01:45 and 02:15 UT, for different energy ranges (green: 12-25 keV, cyan: 25-50 keV, yellow: 50-100 keV) and GOES flux between 1.0 and 8.0 \AA \ (dashed line). The vertical dashed-dotted lines at 01:48 and 02:00 UT represent the time of the two magnetic maps used in the study, and the five vertical dotted lines represent the mean times when X-ray images are produced.}
\label{rhessi_counts}
\end{figure}

\begin{figure}
\begin{center}
\includegraphics[width=\linewidth]{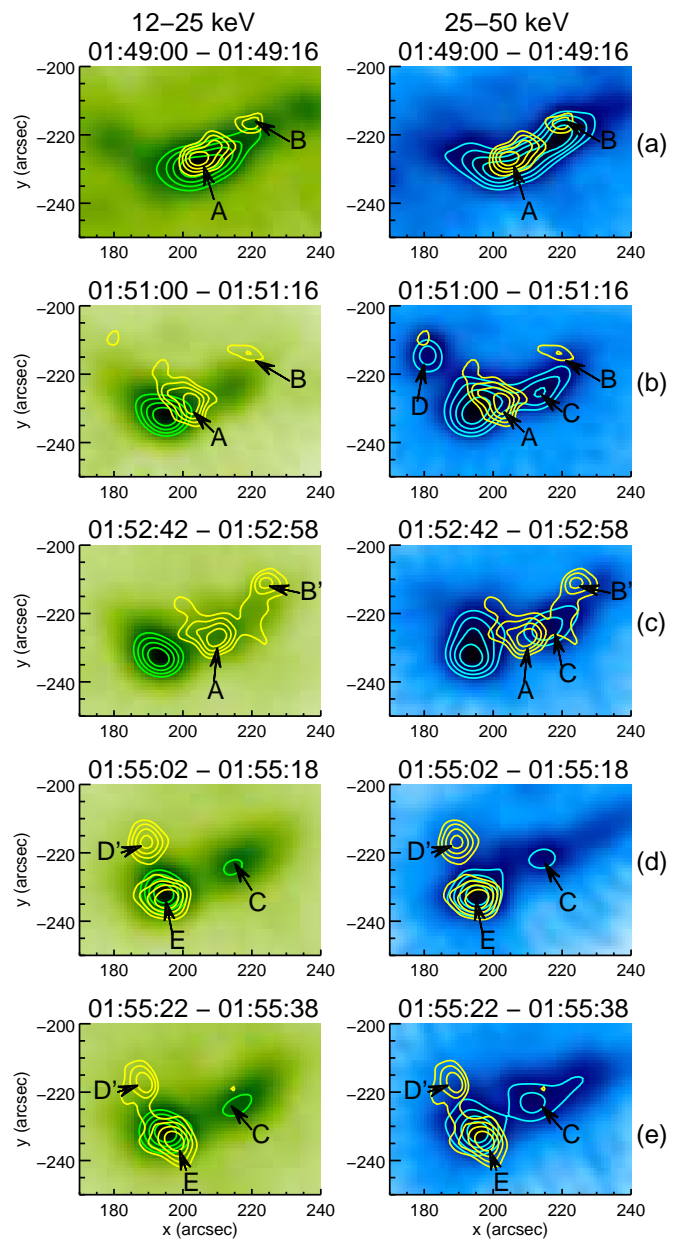}

\end{center}
\caption{Left: RHESSI images at 12-25 keV (green), with contours at 12-25 keV (green) and 50-100 keV (yellow) overlaid on the images. Right: RHESSI images at 25-50 keV (blue), with contours at 25-50 keV (blue) and 50-100 keV (yellow) overlaid on the images. Images and contours are shown for five time intervals integrated over 16 seconds (from top to bottom: 01:49:00-01:49:16, 01:51:00-01:51:16, 01:52:42-01:52:58, 01:55:02-01:55:18 and 01:55:22-01:55:38 UT) corresponding to some peaks in the 50-100 keV range. Images are obtained using detectors 2F, 3F, 4F, 5F, 6F, 7F, 8F and 9F and the CLEAN algorithm. The green, blue and yellow contours are 60\%, 70\%, 80\% and 90\% of the image maximum value.}
\label{rhessi_images}
\end{figure}

\begin{figure}
\begin{center}
\includegraphics[width=\linewidth]{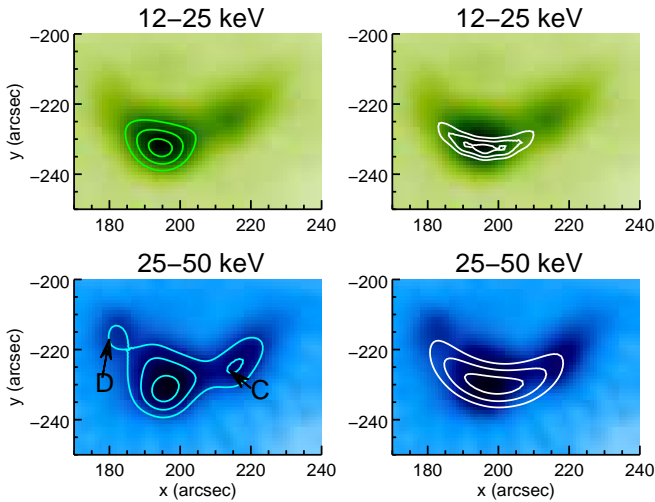}
\end{center}
\caption{RHESSI images and contours at 12-25 keV (green) and 25-50 keV (cyan) obtained between 01:50:48 and 01:51:28 UT using two different methods. Left: CLEAN images and contours at 50\%, 70\%, and 90\% of the maximum. Right: CLEAN images and contours reconstructed with the visibility forward fit technique (50\%, 70\% and 90\% of the maximum).}
\label{visff}
\end{figure}

The X2.2 flare on Feburary 15 2011 was detected by RHESSI in the 3-100 keV energy range.
The RHESSI corrected count rates are presented in figure \ref{rhessi_counts}, together with the X-ray flux from GOES.
In this figure the count rates are corrected from the changes of attenuator state during the flare; indeed, there was no attenuator (attenuator state A0) until around 01:47:10 UT, then one attenuator (A1) was in front of the detectors, and finally at 01:49:40 UT two attenuators (A3) were in place. At 02:08:50 UT, only the first attenuator was left in place.
The peak of the X-ray emission in the 12-25 keV range is around 01:54 UT.

While the temporal evolution of the count rate between 12 and 50 keV is relatively smooth, the emission between 50 and 100 keV is more structured.
Nevertheless the different peaks of emission in the 50-100 keV band have counterparts visible in the 25-50 keV range.
Images created in the 12-25 keV, 25-50 keV and 50-100 keV energy bands, for different peaks of the 50-100 keV count rate, are shown in figure \ref{rhessi_images}. We used an integration time of 16 seconds, and detectors 2F to 9F.
The first set of images (first time interval) is in the attenuator state A1, whereas the other images are in the attenuator state A3.

\setlength{\tabcolsep}{0.1cm} % cette commande permet de choisir l'espacement entre les colonnes d'un tableau
\renewcommand{\arraystretch}{1.3} % et la c'est pour l'espacement entre lignes

Figure \ref{rhessi_images} shows that the X-ray emission sources both in the 12-25 keV and 25-50 keV energy ranges (in green and blue) have an elongated shape (more than 30 arcsec long) with several local maxima in space (the two main maxima are clearly visible in the images at 25-50 keV for the last four time intervals in figure \ref{rhessi_images}). The 12-25 and 25-50 keV emissions keep this elongated form during most of the flare. Moreover, the 25-50 keV emission is more extended in size than the 12-25 keV one, in each time interval.
This has been verified first by measuring the length of the X-ray sources using the 50 \% contours in the CLEAN images, as illustrated in figure \ref{visff} (left).
To better quantify the size of the X-ray source, the visibility forward fitting technique was further used (see \citealt{schmahl_et_al_2007} for the definition of visibilities and \citealt{xu_et_al_2008} for examples of application). This algorithm allows to compute not only the size of the source but also the error on the size once a model has been chosen.
In order to get sufficient statistics, images were created with a 40 seconds integration time (instead of the 16 seconds of figure \ref{rhessi_images}).
The result of the visibility forward fit reconstruction using a curved elliptical Gaussian loop \citep{xu_et_al_2008} as a model is shown in figure \ref{visff} (right column) for the interval 01:50:48 UT - 01:51:28 UT centered around interval (b).
\begin{table}
\caption[10pt]{Loop lengths (in arcseconds) inferred from the visibility forward fit algorithm for 40 second time intervals centered around (b) and (c).}
\label{sizes}
\begin{center}
\begin{tabular}{@{}ccccc@{}}
\hline
   & 12-18 keV & 18-26 keV & 26-40 keV & 40-60 keV \\
\hline
(b) & $28.8 \pm 1.4$ & $32.6 \pm 1.0$ & $40.4 \pm 1.8$ & $44.5 \pm 4.3$ \\
(c) & $28.8 \pm 0.85$ & $29.4 \pm 1.3$ & $46.3 \pm 1.9$ & $48.5 \pm 3.4$ \\
\hline
\end{tabular}
\end{center}
\end{table}
To better determine the evolution of the length of the X-ray source with energy, images were built in narrower energy bins. It is finally found that for the intervals centered around (b) and (c), the length of the source is increasing logarithmically with energy with a slope of $0.23 \pm 0.03$ and $0.60 \pm 0.20$ (see table \ref{sizes}), as already observed in some flares \citep{xu_et_al_2008}.

At higher energies (50-100 keV) the X-ray emission comes from compact sources: A and B visible in intervals (a), (b) and (c) and sources D and E in intervals (d) and (e).
In intervals (a) to (c), sources A, B and B' are localized along the elongated structure seen at lower energies.
While sources B and B' seem to be located at a footpoint of the elongated sources observed at 12-50 keV, source A is located close to the middle of this structure. Between intervals (a) and (c), it is noticeable that source B has moved of about 9 arcseconds to the north-west (source B'), while source A has itself moved of around 3 arcseconds towards the west.

A more noticeable change of configuration of the 50-100 keV sources occurs between intervals (c) and (d) when sources A and B/B' disappear and two new sources, E and D' appear. The two sources furthermore seem to be footpoints of a loop perpendicular to the elongated structure seen at 12-50 keV. It can be noted than source E at 50-100 keV is cospatial with some local maximum emission at 12-25 and 25-50 keV and that at the same position of source D', a source D also appeared in interval (b), but in the 25-50 keV energy range.

The spectral analysis (performed using the two functions vth and thick2 in the RHESSI spectral analysis software) shows that the emission  is a combination of a thermal component with a temperature in the 25-35 $\times 10^{6}$ K range and of a non-thermal component produced by energetic electrons. There is a hardening of the non-thermal electron spectrum starting from interval (d), as the electron spectrum spectral index $\delta$ evolves from [-7,-6.5] in intervals (a), (b) and (c) to [-5,-4.6] in intervals (d) and (e).
For the time intervals of the images shown in figure \ref{rhessi_images}, the spectral analysis shows that while most of the emission in the 12-25 keV range is of thermal origin, the emission in the 25-50 keV range is mostly produced by non-thermal electrons (the proportion of non-thermal emission in this energy range is greater than 75\% in all images).

\begin{figure}
\begin{center}
\includegraphics[width=0.9\linewidth]{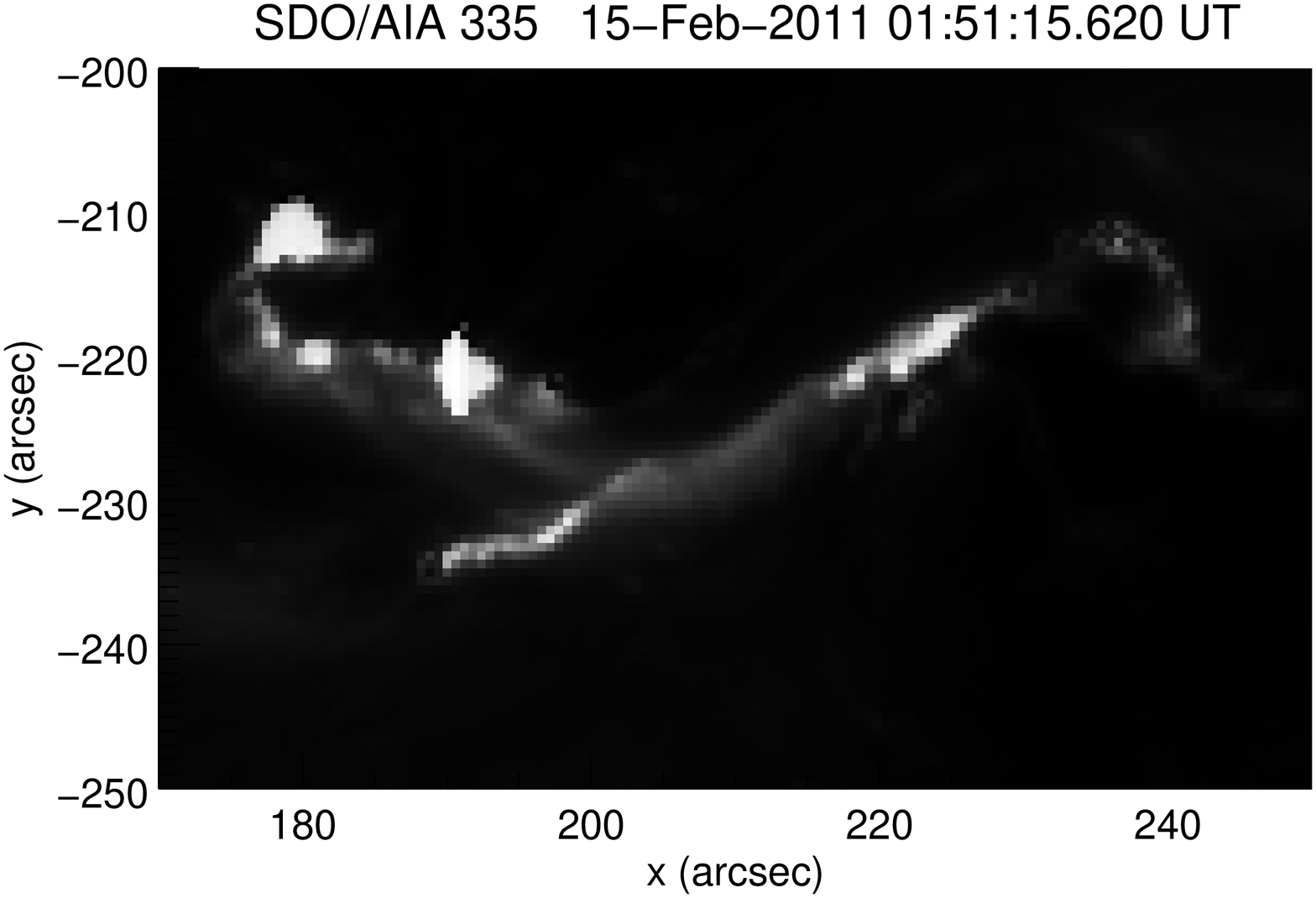}\\
\includegraphics[width=0.9\linewidth]{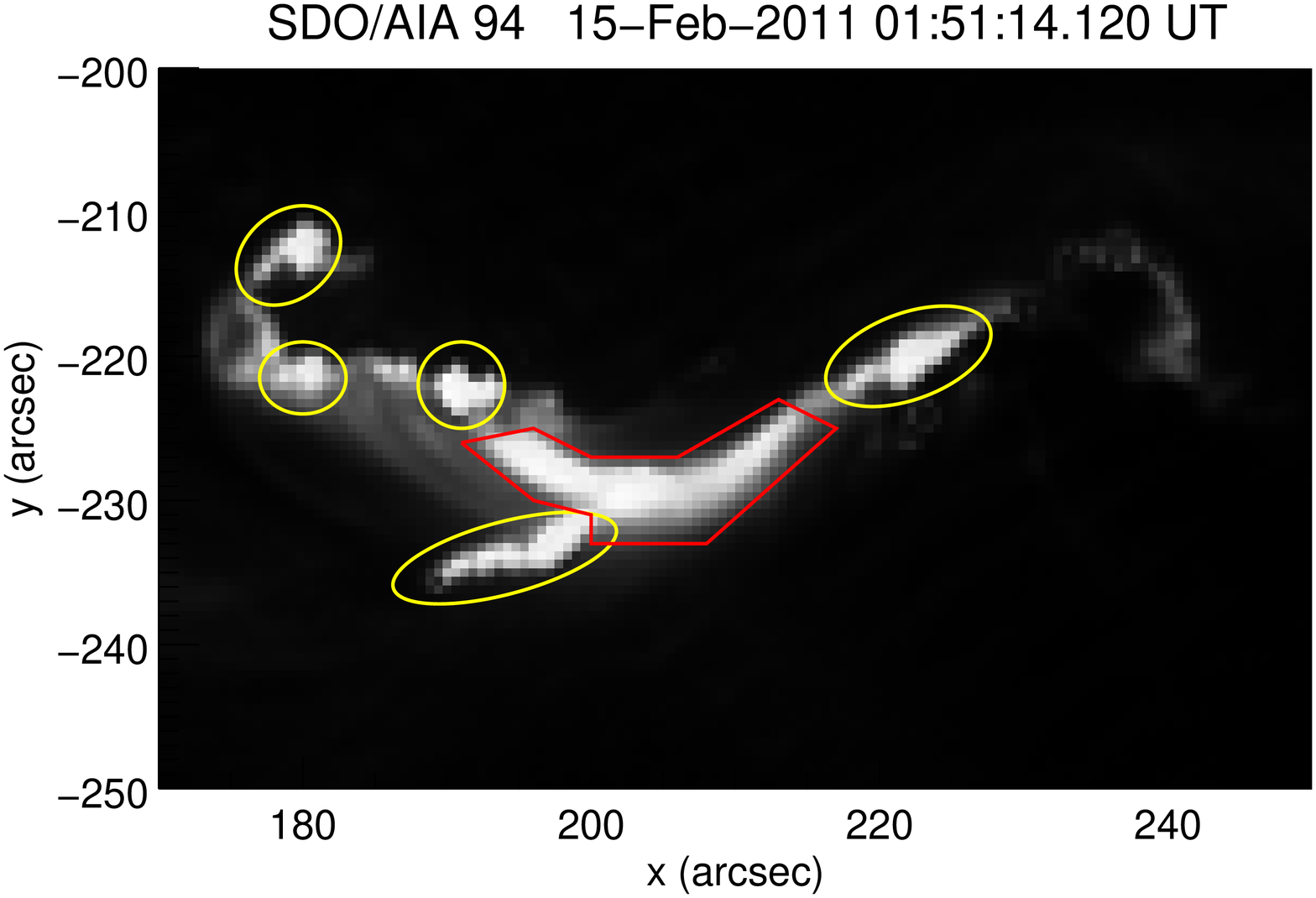}\\
\includegraphics[width=0.9\linewidth]{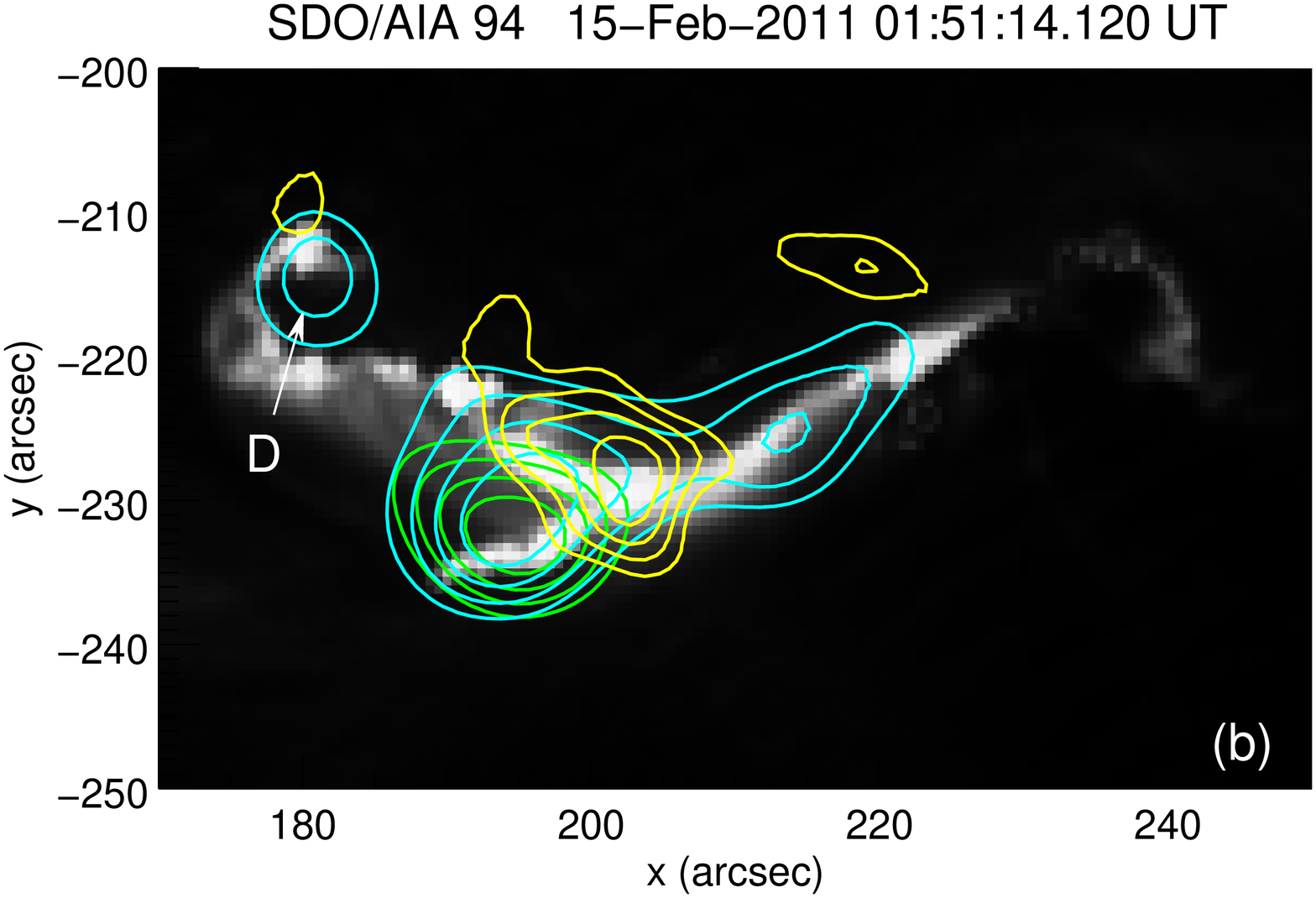}
\end{center}
\caption{SDO/AIA image at 335 \AA \ (top), and 94 \AA \ (middle and bottom), on February 15 2011, around 01:51:15 UT. In the middle image, the region bordered in red is not clearly detected at 335 \AA \ (most probably the plasma temperature is higher than 6 $\times 10^{6}$ K; see text); regions bordered in yellow are seen in both channels (probably the plasma temperature is close to 1 $\times 10^{6}$ K). Bottom: image at 94 \AA \ on which RHESSI contours are overlaid. The green, blue and yellow contours are the RHESSI emissions at 12-25 keV, 25-50 keV and 50-100 keV respectively, integrated between 01:51:00 and 01:51:16 UT. The contour levels are similar to the ones in figure \ref{rhessi_images}.}
\label{temperature}
\end{figure}

\begin{figure}
\begin{center}
\includegraphics[width=0.9\linewidth]{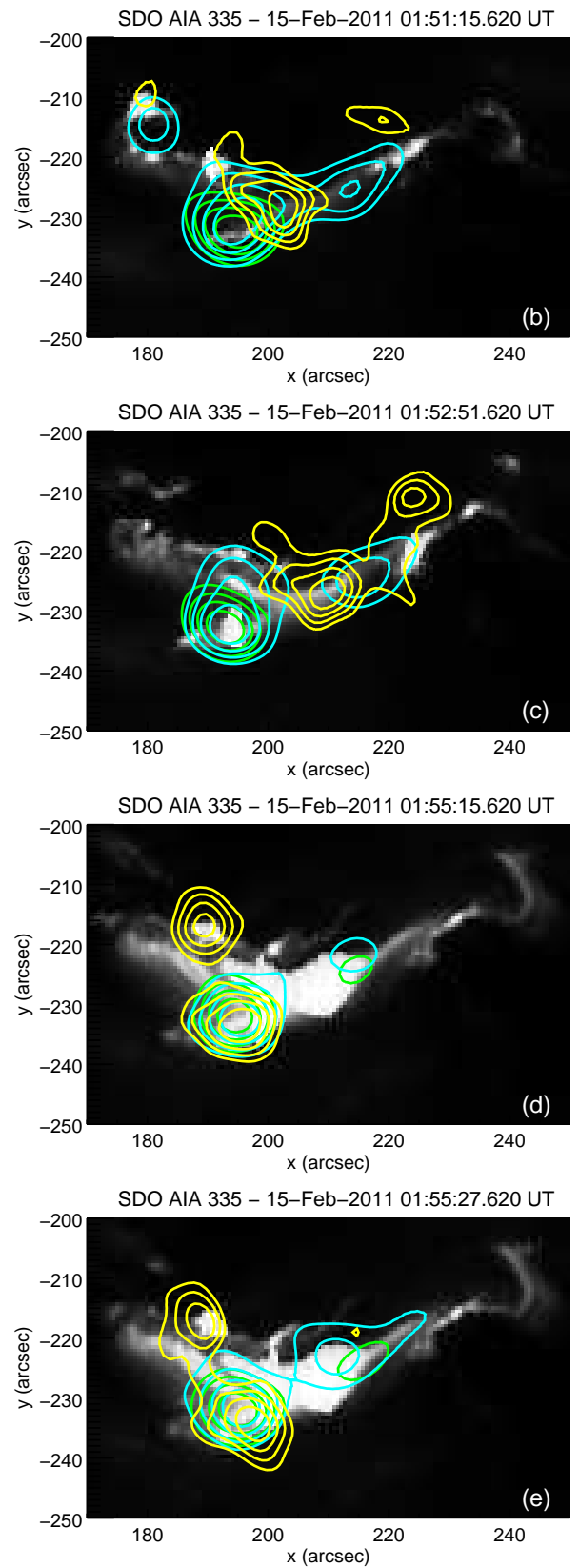}
\end{center}
\caption{SDO/AIA images at 335 \AA \ (grey scale), of a part of the active region 11158, on February 15 2011. RHESSI X-ray contours 12-25 keV (green), 25-50 keV (cyan) and 50-100 keV (yellow) at 01:51:08, 01:52:10, 01:55:10 and 01:55:30 UT are overlaid (contour levels are the same as in figure \ref{rhessi_images}).}
\label{335A}
\end{figure}

\subsection{Comparison with the EUV data from SDO/AIA}

Emissions in both 94 \AA \ and 335 \AA \ channels at around 01:51 UT are shown in figure \ref{temperature} (middle and top pannels respectively).
While the 94 \AA \ emission shows an elongated structure together with more compact sources at the edge, at 335 \AA \ most of the emission arises from the compact sources.
Figure \ref{temperature} (middle) furthermore shows that some structures are emitting in both 94 \AA \ and 335 \AA \ channels (yellow contours), whereas other are emitting only at 94 \AA \ (red contours). This can be interpreted considering the different channel response to source temperature. Table 1 from \cite{aia} gives the characteristic emission temperatures of the plasma for the 94 \AA \ and 335 \AA \ channels: $6.3 \times 10^{6}$ K and  $2.5 \times 10^{6}$ K respectively.
However, as shown in figure 13 of \cite{aia}, the different channels are sensitive to a wide distribution of temperatures.
From this figure, it is deduced that, if a structure is bright at 94 \AA \ and 335 \AA , the plasma temperature is most probably around $1 \times 10^{6}$ K (yellow contours), while if it is bright only in the 94 \AA \ channel, it would be much higher (more than $6 \times 10^{6}$ K, red contours). The superposition of the elongated part of the X-ray sources below 50 keV with the elongated structure seen at 94 \AA \  in figure \ref{temperature} (bottom) strongly supports this interpretation since the spectral X-ray analysis shows that the X-ray emission below 50 keV is partly emitted by a hot plasma with a temperature in the range 25-35 $\times 10^{6}$ K.
Figure \ref{temperature} (bottom) also shows that source D in image (b) in figure \ref{rhessi_images} is roughly co-spatial with a EUV source detected both at 335 and 94 \AA . This source is most probably the footpoint of coronal loops in which energetic electrons are injected.
Figure \ref{335A} shows the evolution with time of both X-ray and EUV sources at 335 \AA . Only the 335 \AA \ can be used for this comparison since the emission is saturated at 94 \AA \ after 01:52 UT. In the last two images (intervals d and e) in which the configuration of HXR sources at 50-100 keV is dramatically changed compared to interval (c), it should be noted that  new arcade-like structures appeared in EUV between the two 50-100 keV sources.

\subsection{Comparison with the vertical electric current densities}
\label{comparaison_courants}

\begin{figure}[ht]
\begin{center}
\includegraphics[width=0.83\linewidth]{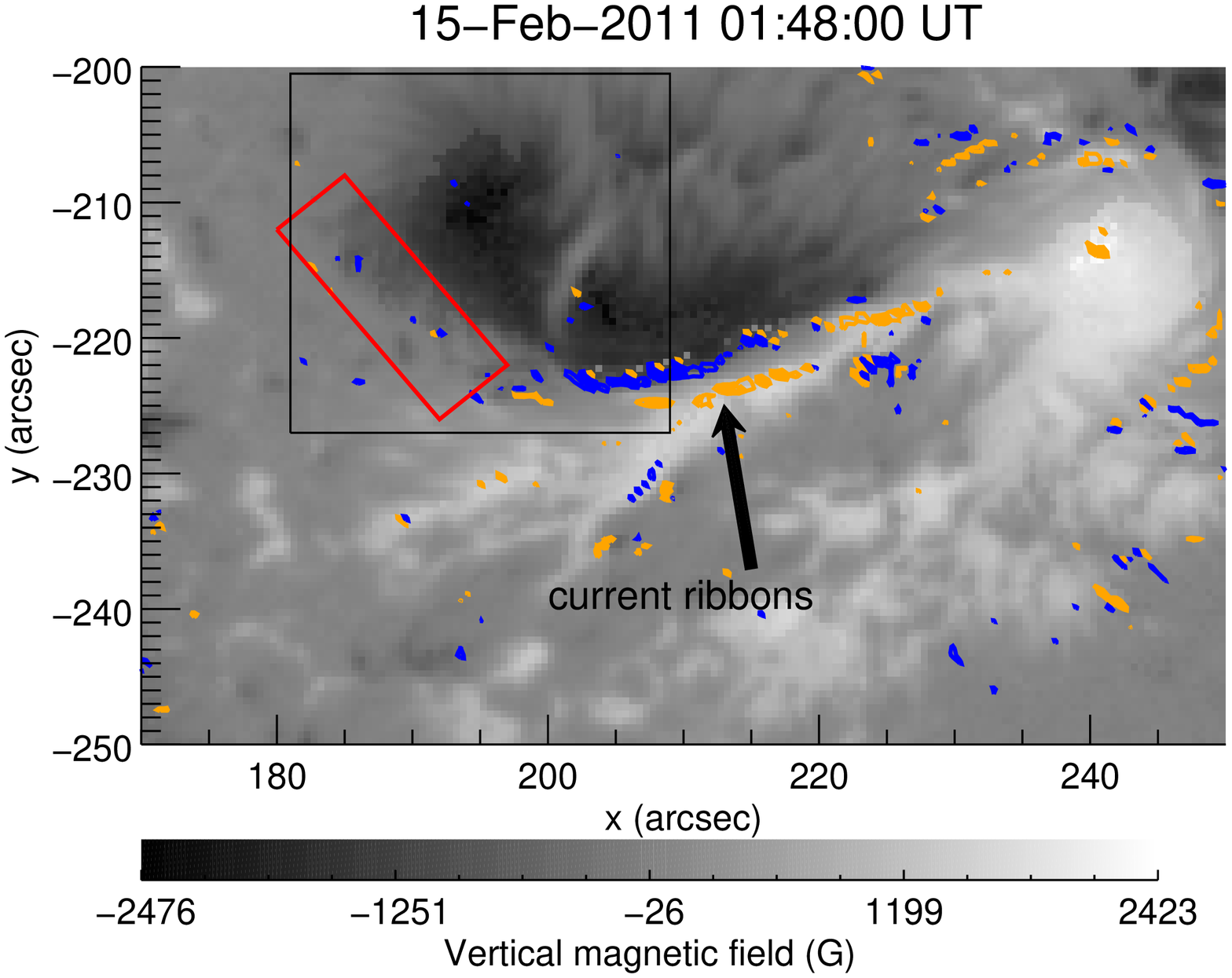}\\
\includegraphics[width=0.83\linewidth]{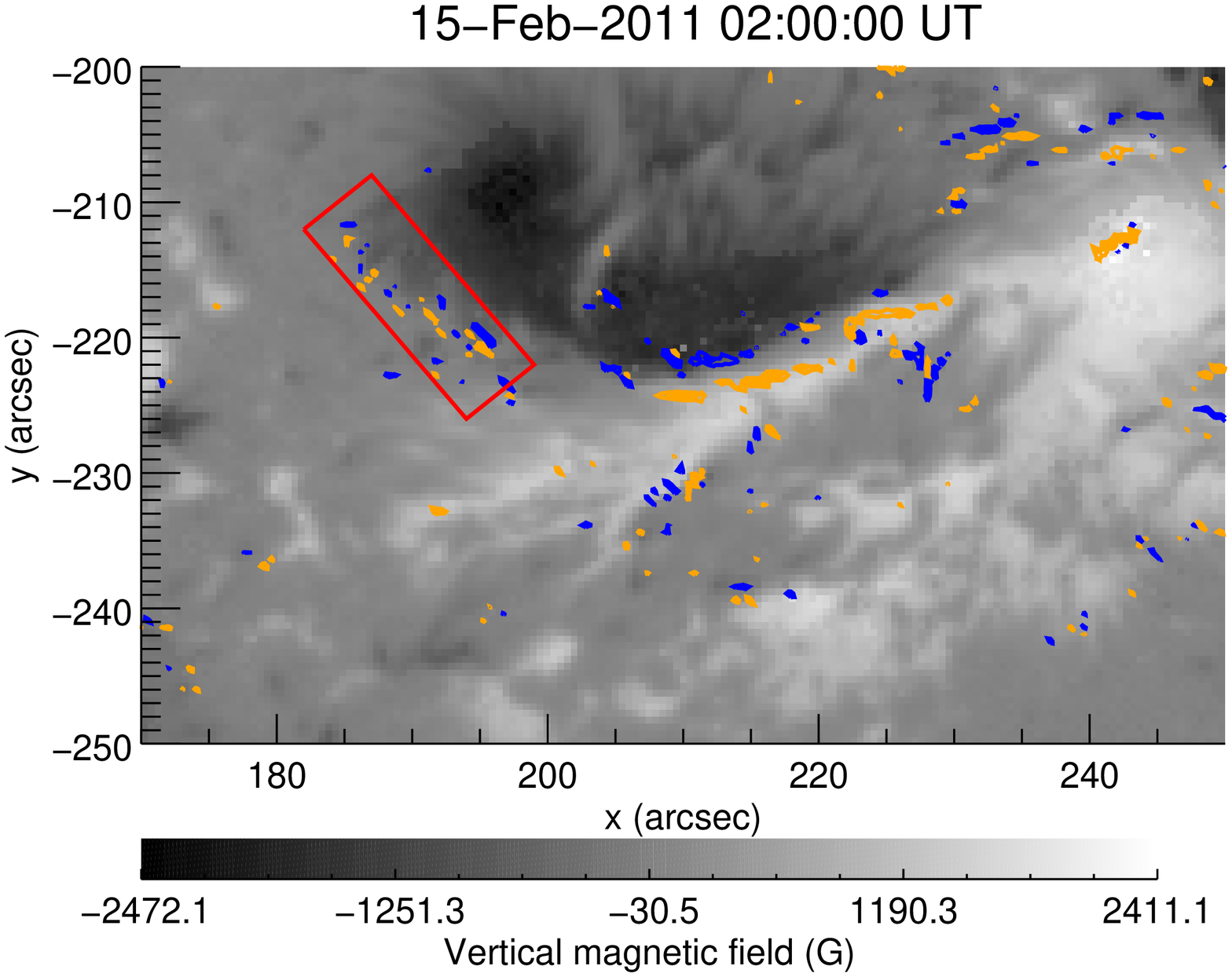}
\end{center}
\caption{Magnetic field maps (in grey scale) from SDO/HMI of a part of the active region 11158, on February 15 2011, at 01:48:00 UT (top) and 02:00:00 UT (bottom), represented in the plane-of-sky. The orange and blue contours represent the positive and negative vertical electric current densities respectively, with the magnitude greater than 100 mA/m$^{2}$. The black arrow in the top pannel indicates the current ribbons extending from 195 arcsec to 230 arcsec in the x-direction, and laying between -220 arcsec and -230 arcsec in the y-direction. The red box enlightens the primary difference between the two maps : in this box, the total negative vertical current density varies from $-974.9 \times 10^{9}$ A to $-1062 \times 10^{9}$ A, and the total positive vertical current density varies from $+616.9 \times 10^{9}$ A to $+762.4 \times 10^{9}$ A. Therefore, the total increase of vertical current density is of 232 $\times 10^{9}$ A for the area of the red box. This represents an increase of 15\% of the total current density in this area.} %  (this correspond to a mean increase of 4.5 mA/m$^{2}$ per pixel.}
\label{magnetic}
\end{figure}

Figure \ref{magnetic} shows the combination of the vertical magnetic field and of the vertical currents for the field of view indicated by the black box in figure \ref{magn_map}, i.e. the part where the X-ray emissions of the flare come from.
The improved spatial resolution of magnetograms allows to distinguish very fine structures such as the current ribbons \citep{janvier}, which are visible on figure \ref{magn_map} as the alignement of close positive and negative currents and which are shown by the black arrow in the top pannel of figure \ref{magnetic}. These current ribbons are present in both maps at 01:48 and 02:00 UT.
Note that these narrow current ribbons have a fragmented appearance with many knots of high current densities distributed along the ribbons.
The other noticable feature in figure \ref{magnetic} (bottom) is the appearance, in the region marked by the red box, of strong currents. The current density in this region is increased by 15\% between the two maps. This evolution of the currents has already been noted by \cite{janvier}.

\begin{figure*}
\begin{center}
\includegraphics[width=0.4\linewidth]{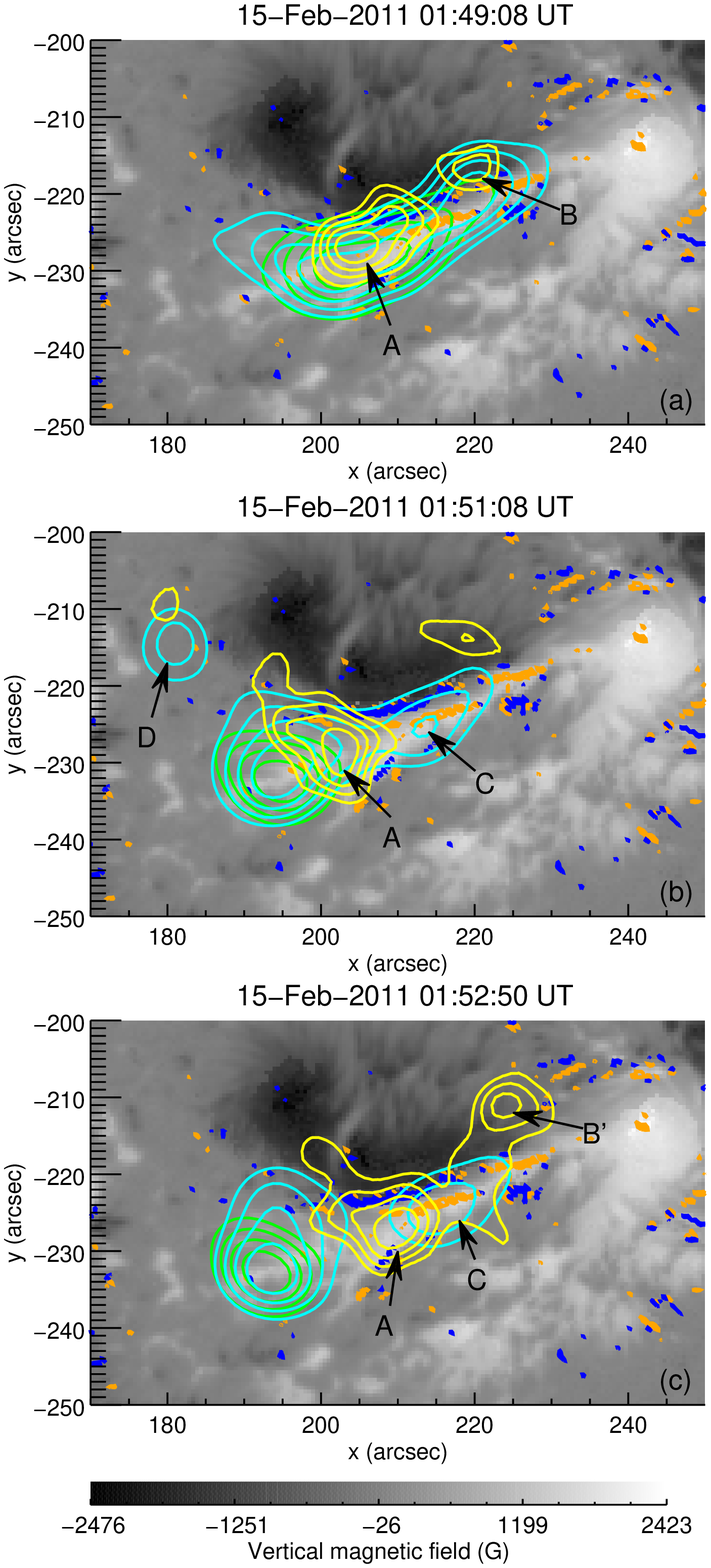}
\includegraphics[width=0.4\linewidth]{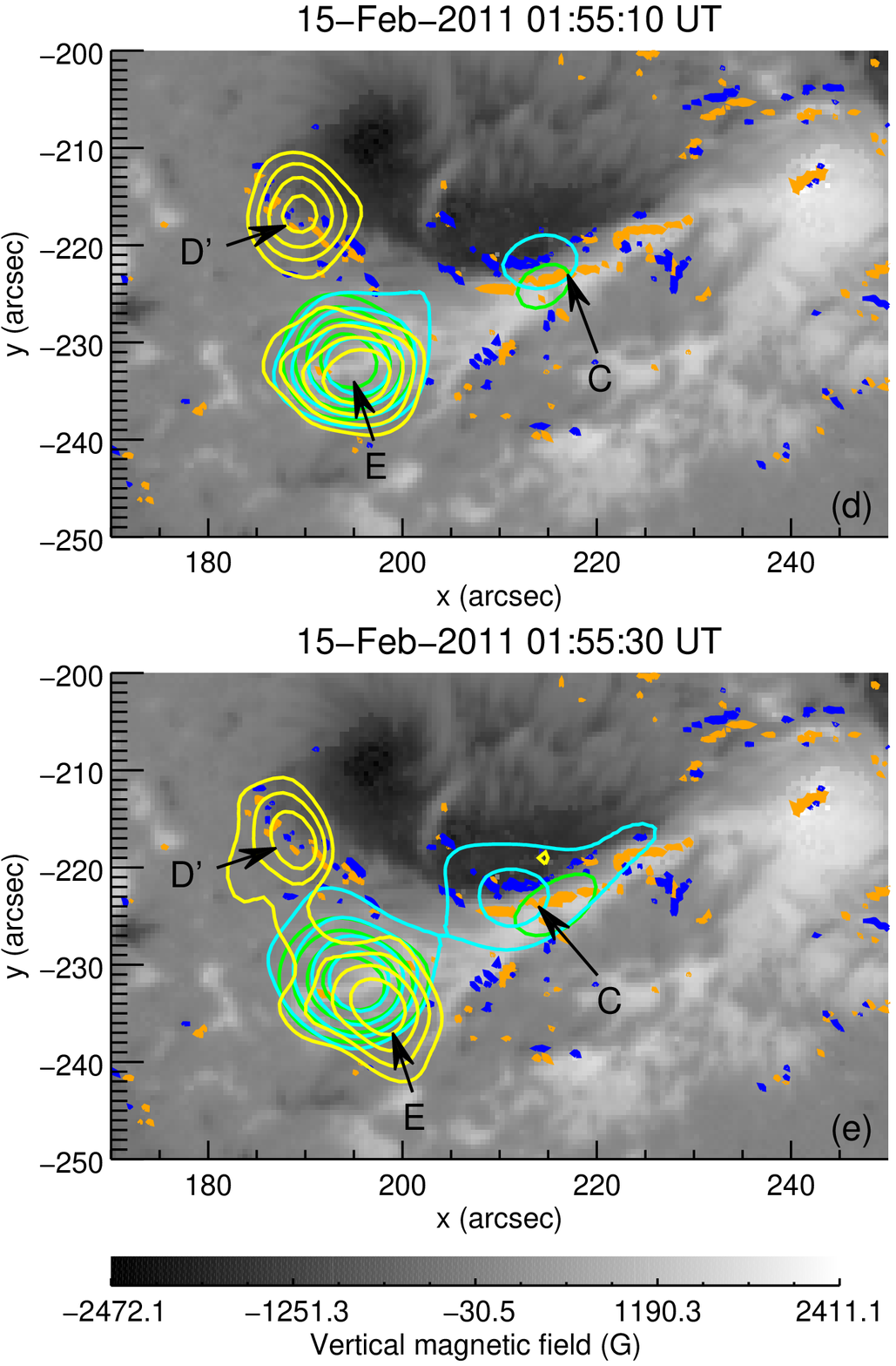}
\end{center}
\caption{Magnetic field maps (in grey scale) from SDO/HMI of a part of the active region 11158, on February 15 2011, at 01:48:00 UT (left) and 02:00:00 UT (right). The orange and blue contours represent the positive and negative vertical electric current densities respectively, with amplitude $>$ 100 $mA/m^{2}$. The green, cyan and yellow contours are the X-ray emissions (from RHESSI) at 12-25 keV, 25-50 keV and 50-100 keV respectively, integrated between (a) 01:49:00 and 01:49:16, (b) 01:51:00 and 01:51:16, (c) 01:52:42 and 01:52:58, (d) 01:55:02 and 01:55:18, and (e) 01:55:22 and 01:55:38 UT. X-rays have been imaged with the algorithm CLEAN, using collimators 2, 3, 4, 5, 6, 7, 8 and 9. The contours corresponds to 60, 70, 80 and 90 \% of the maximum for the X-ray emissions.}
\label{5images}
\end{figure*}

Figure \ref{5images} shows the evolution of X-ray sources with time, overlaid on the vertical magnetic field and current density maps at the photospheric level obtained at respectively 01:48:00 UT (left) and 02:00:00 UT (right).
Two main results can be drawn from this figure:
\begin{itemize}
\item Part of the elongated X-ray sources observed between 12 and 50 keV in intervals (a),(b),(c) overlay the current ribbons.
\item The new X-ray source D' appearing at 50-100 keV in intervals (d) and (e) is located in the region where new vertical photospheric electric currents appeared between 01:48 and 02:00 UT.
\end{itemize}

\begin{figure}
\begin{center}
\includegraphics[width=0.7\linewidth]{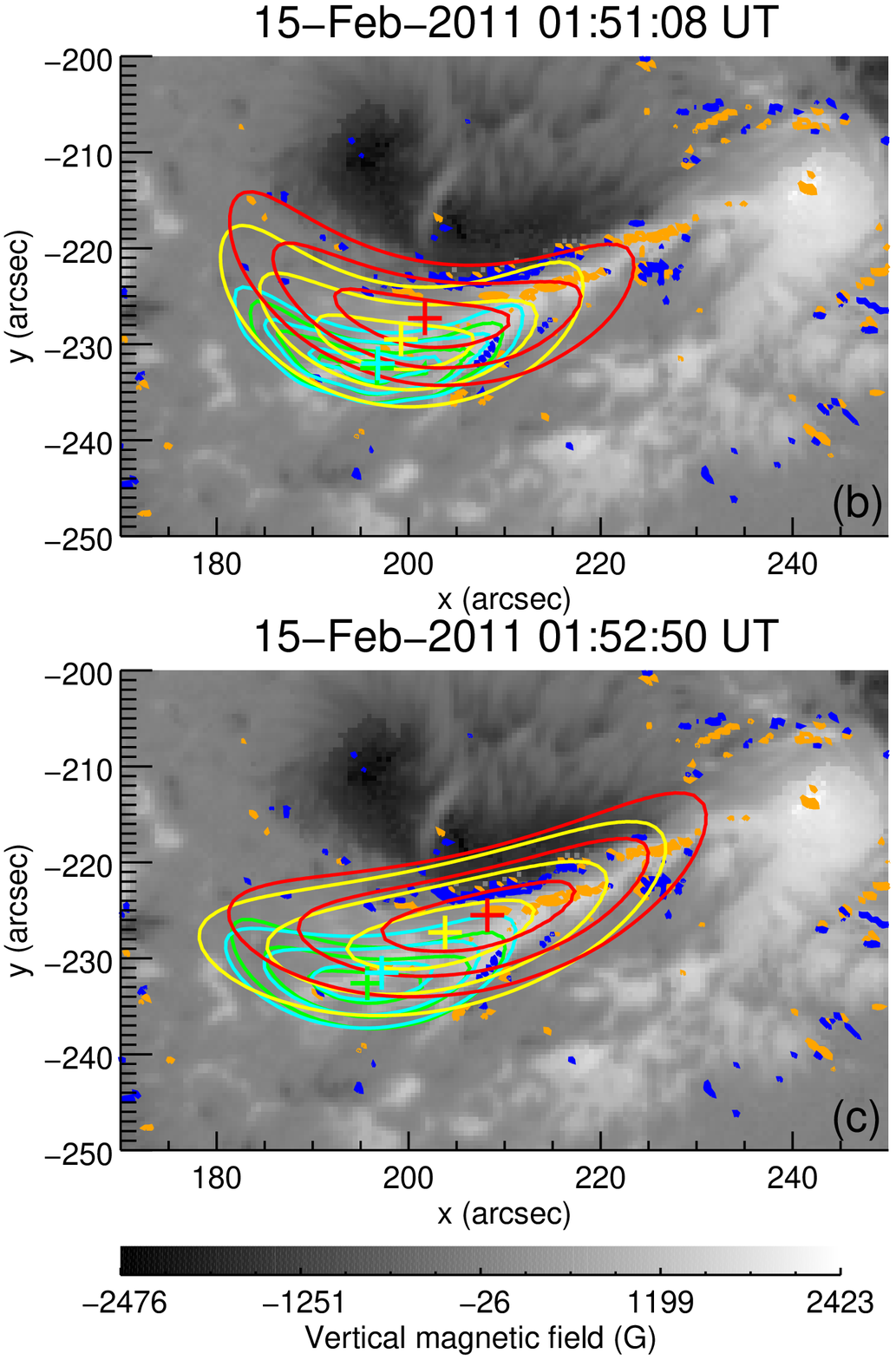}
\end{center}
\caption{Magnetic field map (in grey scale) from SDO/HMI of a part of the active region 11158, on February 15 2011, at 01:48:00 UT. The orange and blue contours represent the positive and negative vertical electric current densities respectively, with amplitude $>$ 100 $mA/m^{2}$. The green, cyan, yellow and red contours are the X-ray emission (from RHESSI) at 12-18 keV, 18-26 keV, 26-40 keV and 40-60 keV respectively, integrated between (b) 01:50:48 and 01:51:28 and (c) 01:52:30 and 01:53:10 UT. X-rays have been imaged with the Visibility Forward Fit technique using collimators 2, 3, 4, 5, 6, 7, 8 and 9. The contours corresponds to 50, 70 and 90 \% of the maximum for the X-ray emissions. The crosses are at the location of the maxima of the sources.}
\label{vff_superposition}
\end{figure}

The distance of the centroid of each X-ray source seen in Figure \ref{5images} to the nearest current concentration has been estimated.
Most of the HXR centroids are found to be within a distance lower than 4 arcseconds to the closest part of a current ribbon. However, two sources make exception: the centroid of source E (see images (d) and (e)) which is  distant of 8 to 12 arcsec to the nearest current ribbon, and the centroid of source B on image (b) which is at 6 arcsec to the nearest current ribbon.
Figure \ref{vff_superposition} shows the X-ray images reconstructed with the visibility forward fit technique superimposed on the magnetic field and current density maps for intervals (b) and (c). The distance of the maximum of the elongated source (in the narrow energy ranges used in section \ref{obs_hxr}, see e.g. table \ref{sizes}) to the nearest current concentration has also been estimated. This distance is found to decrease with photon energies, being of the order of 8 to 10 arcseconds for energies below 26 keV and below 6 arcsec for energies above 26 keV.

%======================================================================= 4 = DISCUSSION - INTERPRETATION ===============================================================================================

\section{Discussion and interpretation}
\label{section4}

\subsection{How to interpret the spatial configuration of X-ray and EUV sources?}

During the flare, the X-ray sources below 50 keV have an elongated shape (more than 30 arcseconds in length), cospatial with the EUV emission at 94 \AA. A large proportion of the X-ray emission in the 25-50 keV range is however mostly non-thermal (more than 75\% of the total emission in this energy range). It can be furthermore noted that the length of the source is systematically increasing with energy from 12-18 keV to 40-60 keV (see table \ref{sizes}).

Such an elongated shape of the hard X-ray emission as well as the increase of the source size with energy is not the standard situation. However, a few events have been reported in which non-thermal HXR emissions at 25-50 keV had an elongated shape resulting from thick-target interactions in a dense coronal loop (see e.g. \citealt{veronig_brown_2004}).
The evolution with energy of hard X-ray source sizes was furthermore estimated by \cite{xu_et_al_2008} for several limb events on extended sources. They found, as in the present case, a slow increase of the source size with energy (the logarithmic slope found for several events is in the range $0.11 \pm 0.04$ - $0.76 \pm 0.03$). They also predicted the evolution of the source size with energy when X-ray emission results from a thermal population of electrons produced at the apex of the loop or when X-ray emission results from non-thermal electrons accelerated in an extended region at the loop apex. They found that in the case of thermal emissions, the source size should decrease with energy, while it should increase in the case of non-thermal emissions. Combining these predictions with the present observations of X-ray source sizes (systematic increase with energy from 12-18 keV to 40-60 keV), we therefore conclude that in this flare the X-ray emissions in the 25-50 keV range coming from the elongated sources arise primarily from non-thermal emissions (in agreement with the spectral analysis of section \ref{obs_hxr}).
Furthermore following \cite{xu_et_al_2008} for the interpretation of the evolution of the source size with photon energy, it can be deduced that the acceleration region is extended: the size is found to be around 28 arcseconds and the density of the loop of the order of $3 \times 10^{10} cm^{-3}$. Such values are in good agreement with what has been found for a few events by \cite{kontar_et_al_2011} and \cite{guo_et_al_2012}. In their work, \cite{xu_et_al_2008} modelled the transport of electrons on the basis of a standard collisional transport model. It was later shown by \cite{kontar_et_al_2014} that an increase of X-ray source size with energy can also be predicted in the context of a diffusive transport model of electrons, in which the turbulent pitch angle scattering of energetic electrons on magnetic fluctuations leads to an enhancement of the coronal HXR source relative to the footpoints. Comparing the predictions of this last model with our observations, the size of the acceleration region is found to be 22 arcsec ($1.6 \times 10^{4}$ km) and the electron mean free path of the order of $10^{9}$ cm, which is in agreement with the results of \cite{kontar_et_al_2014} for another flare.

At higher energies (>50 keV), the X-ray emission is produced by a non-thermal population of electrons and the emissions come from more compact sources: sources A, B and B' in images (a) to (c) and sources D' and E in images (d) and (e). While sources B and B' appear to trace the impact of energetic electrons at the footpoint structures, the 50-100 keV source labelled A in images (a), (b) and (c) is located in the middle of the elongated structure observed at lower energies. Its location is also very close to the position of the maximum of the elongated X-ray source in the 40-60 keV range. This high-energy source could be similar to coronal sources more easily observed in the case of limb events \citep{krucker_et_al_2008,krucker_et_al_2010,chen_and_petrosian_2012,su_et_al_2013}, but projected on the disk in our case.
The combined location of coronal and loop-top sources could trace the location where the main reconnection process occurs i.e. above the loop in which electrons then  propagate \citep[see e.g.][]{su_et_al_2013}. In this context, the evolution of the location of sources A, B and B' between time intervals (a) and (c) most probably follow, the evolution of the acceleration site.
It must be recalled that movement of X-ray footpoints are usually observed in the course of flares and related to the magnetic reconnection process \citep[see e.g.][for a review]{fletcher_2011}. A more significant change in the configuration of HXR sources above 50 keV is observed later in the flare (images (d) and (e)).
Simultaneously with the hardening of the non-thermal electron spectra, new high energy HXR sources appear (sources D' and E) most probably located at the footpoints of an arcade of loops which appeared in EUV at the same time.
These observations show that after $\approx$ 01h55 UT, magnetic reconnection and the resulting plasma heating and particle acceleration occurs in a new structure.
It is finally noticeable that the position and orientation of the new EUV bright structure are consistent with the ones of the post-flare loops appearing in the 3D MHD simulation of the flare performed by \cite{inoue_2014}.

%----------

\subsection{How to interpret the relation between X-ray emissions and photospheric vertical electric currents?}

\begin{figure*}
\begin{center}
\includegraphics[width=0.9\linewidth]{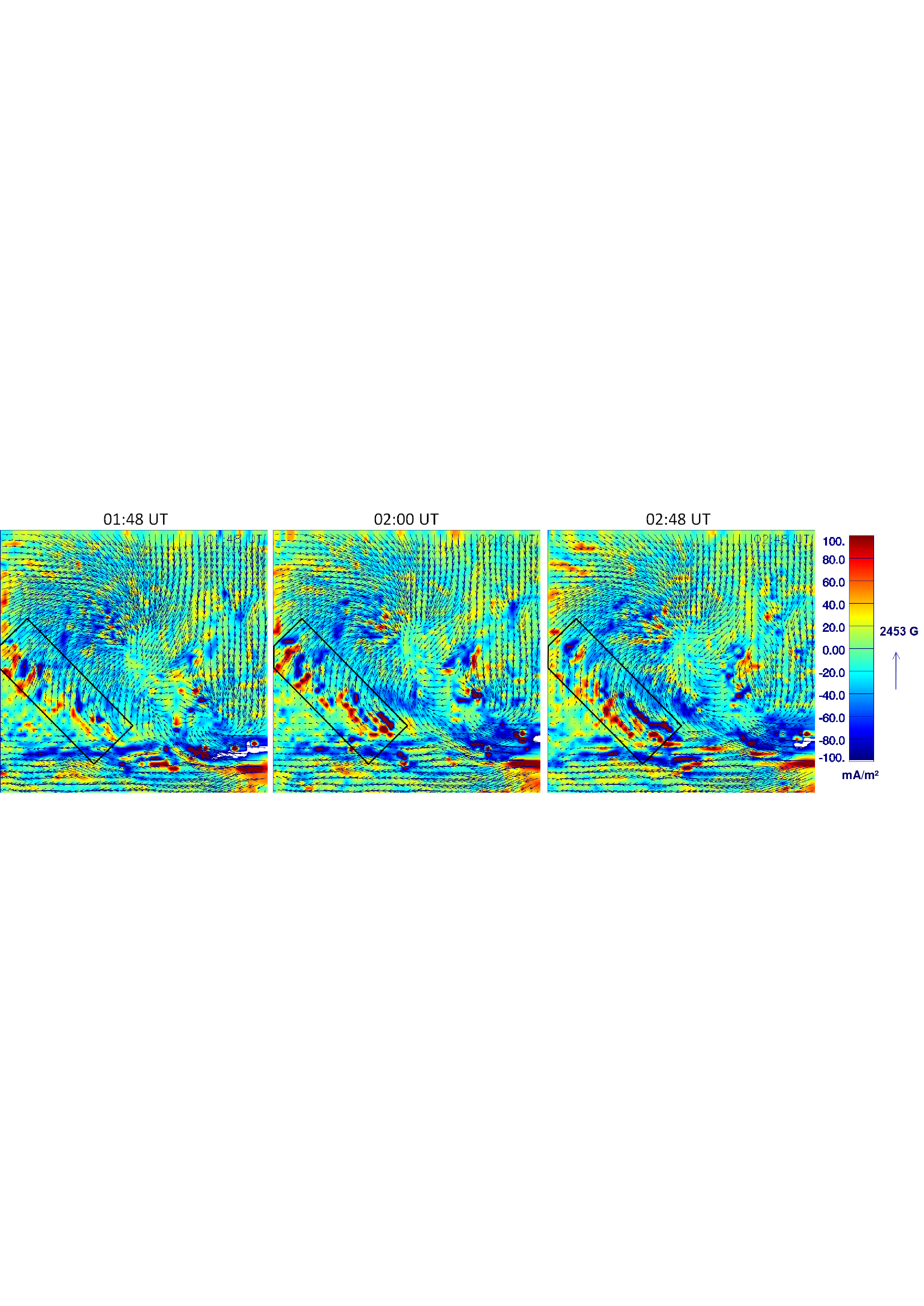}
\end{center}
\caption{Vertical current density (color) and horizontal magnetic field (arrows, see scale), for the region indicated by the black box in figure \ref{magnetic} ($28''\times26''$ centered at x$=$195 arcsec and y$=$-214 arcsec), at 01:48 UT (left), 02:00 UT (middle) and 02:48 UT (right). The region where the current density increases by 15\% between 01:48 and 02:00 UT is surrounded by the black rectangle on the three present figures.}
\label{zoom}
\end{figure*}

As discussed in section \ref{comparaison_courants}, part of the elongated thermal and non-thermal X-ray sources (observed at energies below 50 keV) are located above the narrow and elongated current ribbons measured at the photospheric layer. These elongated X-ray structures have furthermore been identified in the previous section as being produced in the corona by electrons injected from an extended acceleration region located close to the apex of the magnetic structures. At energies above 50 keV, the X-ray source A is furthermore located very close (in projection) to the main current ribbon (shown by a black arrow in the top panel of Figure \ref{magnetic}) revealing the close association between the location of the acceleration region and the location of the electric current sheets.

As it has been discussed in \cite{janvier}, the photospheric current ribbons (closely related to the flare EUV ribbons) are the tracers at the photospheric boundary of the electric current sheets present in the coronal volume. These current layers are themselves formed in regions of strong gradients of magnetic connectivity (see figure 7 of \citealt{janvier} and \citealt{zhao_2014} for the computation of QSLs in the region)
known to be the preferred locations where reconnection can occur \citep[see e.g.][]{demoulin_1996}. As a consequence, plasma heating and particle acceleration can be produced in these regions.

This explains why part of the thermal and non-thermal X-ray emissions produced in the coronal elongated sources are found above the photospheric current ribbons which trace the footprints of the coronal current layers.

At energies above 50 keV, the apparition of the X-ray source D' is accompanied with the appearance of cospatial photospheric currents in the same time interval. Note that the observation of the current density evolution has to be carefully examined since polarization measurements during flares can be biaised, and some of the observed features could result from polarization artifacts due to impact of non-thermal particle beams \citep[see][and references therein]{henoux2013} or induced by resonant scattering due to radiation anisotropies at the edges of flare ribbons \citep{stepan2013}.
\cite{janvier} discussed this issue for this specific event. They analysed cautiously the vector magnetic field in the active region and concluded that the changes in the electric current density are not artifacts due to polarization effects. They argue that the horizontal fields (from which vertical electric currents are derived) are consistent in both maps at 01:36 UT and 02:00 UT with a signal free of artifacts given the smooth rotation of the magnetic field in this region. Figure \ref{zoom} \citep[inspired from figure 4 in][]{janvier} clearly shows a coherent direction of the horizontal magnetic field which varies smoothly within the field of view.
Moreover, the curvature of the horizontal field lines increases between 01:48 and 02:00 UT, which is responsible for the increase of the observed vertical current \citep[see also the observations by][]{petrie_2013}. Finally, the increase in the current density shown in figure \ref{magnetic} is persistent, as it can be seen in figure \ref{zoom}. This is a very strong argument against artifacts due to energetic particle precipitation or flare radiation anisotropies, because at 02:48 UT the GOES X-ray flare has largely decayed. We thus reach the same conclusion as \cite{janvier}: the increase of the photospheric currents in the black box of figure \ref{zoom} is not due to a polarization artifact but reflects real changes in the horizontal magnetic field and is thus real.

As a conclusion, the appearance of the new HXR source D' and the increase in the same region and in the same time interval of the photospheric currents are linked.
The increase in the photospheric currents can be interpreted as a response at the photospheric layer of the change of magnetic topology and current systems in the corona induced by reconnection \citep[see][]{janvier}.
As particle acceleration is also a consequence of the magnetic reconnection in the corona,
it naturally explains the close temporal and spatial association between locally enhanced X-ray emission and increase
in the same region and at the same time of photospheric currents: both result from the same phenomenon, namely magnetic reconnection in the localized coronal current sheets.
It should finally be noted that such a related evolution of X-ray sources and electric currents in the course of a flare has not been reported previously, given the low cadence of magnetic field measurements before SDO/HMI. However, even in the present case, it is clear that the cadence to derive magnetic field maps (and current density maps) from HMI (12 minutes) is still too low to be able to track the evolution of electric currents on the time scales relevant to the evolution of X-ray sources (a few tens of seconds).

\subsection{Comparison with previous results}

As recalled in the introduction,
the link between vertical electric currents and electron precipitation sites as diagnosed by HXR emissions at footpoints has been investigated in the 1990s combining YOHKOH/HXT observations and vector magnetograph data from the Mees Solar Observatory \citep{li,canfield_et_al_1992}.
It was found that the electron precipitation sites as detected via the HXR footpoints are not exactly co-spatial with the regions of highest vertical current densities but adjacent to the current channels.
In this new study, using vector magnetograph data (and thus vertical electric maps) obtained from space (with an improved seeing) with a higher spatial resolution in combination with HXR images obtained at the same time and with improved image dynamics and spatial resolution, the link between electron precipitation sites and current concentrations has been revisited.
These new observations confirm some of the previous result that HXR footpoints indicating electron precipitations sites are not exactly cos-patial with the region of highest electric current densities. This is in particular the case in this flare of the footpoint sources B and E which are at a distance of 6 to 12 arcsec to the nearest current ribbons in the present flare.

HXR footpoints can also be found at the same location as the current ribbons themselves.
This is the case of the HXR footpoint D' seen at energies above 50 keV after 01:55 UT which is
cospatial with new current ribbons which appeared themselves between 01:48 and 02:00 UT.
It should be emphasized that the observation of such a combined evolution of HXR footpoints
and electric currents is reported here for the first time. A complete interpretation of this
observation is not within the scope of this paper aimed at observations but would require
a more complete study combining models of 3D reconnection in flares and particle acceleration.

In addition to HXR footpoint sources, a significant part of the HXR emissions observed in this flare arises from an extended coronal source observed up to 60 keV. Such coronal HXR sources were not observed in the events studied by \citep{li,canfield_et_al_1992}. The comparison of coronal HXR sources with electric currents ribbons is then performed for the first time in this study. It is found that part of the thermal and non-thermal coronal sources just overlay the current ribbons. This should be further investigated in other events and interpreted in the context of combined models of 3D reconnection in flares and particle acceleration but this clearly provides support to the models of particle acceleration in reconnecting current sheets \citep[see e.g.][for a review of acceleration models]{Zharkova}.

\section{Summary and conclusion}
\label{section5}

This study is the first detailed comparison between the spatial distribution of photospheric vertical currents and of X-ray emission sites produced by hot plasma and accelerated electrons. For the first time, X-ray images and maps of electric currents (derived from vector magnetograms) were obtained at the same time with an improved time cadence for the observations of vector magnetograms (and derivation of electric current maps) which furthermore enables to study the coupled evolution of the photospheric electric currents and the HXR emission sites in the course of the flare.

The main results obtained from this first study are the following ones:
\begin{itemize}
\item Parts of the thermal and non-thermal X-ray emissions produced in an extended coronal source overlay the elongated narrow current ribbons observed at the photospheric level.
\item A new hard X-ray source at energies above 50 keV appears in the course of the flare in association with an increase of 15\% of the photospheric current at around the same time and the same location. This shows here a clear link between particle acceleration and reconnecting current sheets.
\end{itemize}

These two results can be qualitatively explained in the context of the commonly admitted scenario in which magnetic reconnection occurs at current-carrying QSLs in the corona with part of the energy released transferred to plasma heating and particle acceleration. Since X-ray (and EUV) emissions are signatures of plasma heating and of particle acceleration, and since photospheric currents can trace the footprints of the coronal currents embedded in QLSs, some spatial and temporal correlation can be naturally expect between X-ray (and EUV) emitting sites and photospheric current ribbons.
In addition, the evolution of magnetic reconnection sites in the course of the flare, may lead to two linked consequences: on one hand, plasma heating and particle acceleration are produced at different locations, potentially leading to the appearance of not only new EUV post-flare loops but also of new X-ray sources; on the other hand, the change of magnetic topology due to the evolving reconnection process leads to an increase of the photospheric current densities as the same place and time \citep[see also][]{janvier}. These new results have been obtained so far on only one flare. A similar study should be further extended in the future to other X-class flares, to examine whether this relation between photospheric current ribbons and X-ray sources is systematically observed.This work furthermore shows that the evolution of horizontal magnetic fields and of derived photospheric vertical electric currents can be observed on timescales of 12 minutes during a strong flare. The cadence is however still too low to derive the detailed timing between the evolution of the fields and currents and the evolution of the energetic electron acceleration and interaction sites.
Finally, this kind of studies would deserve future interpretation and modeling: in particular tentatively coupling acceleration processes to 3D MHD models should allow to derive the location of the reconnecting current sheets and of their observed photospheric traces together with the location of HXR sources.

\begin{acknowledgements}
We thank Mikola Gordovskyy, Guillaume Aulanier, Miho Janvier, Anna Massone and Brigitte Schmieder for their useful comments, as well as the RHESSI team for producing free access to data. SDO data are the courtesy of the NASA/SDO AIA and HMI science teams. The authors thank the anonymous referee for his suggestions which led to major improvements to this manuscript. Sophie Musset acknowledges the CNES and the LABEX ESEP (N$^{\circ}$ 2011-LABX-030) for the PhD funding, and thanks the French State and the ANR for their support through the \textquotedblleft Investissements d'avenir\textquotedblright \ programm in the PSL$\ast$ initiative (convention N$^{\circ}$ ANR-10-IDEX-0001-02).
\end{acknowledgements}

\appendix

\section{Spatial superposition of magnetic field, electric current density and emissions from flares}
\label{changeofframe}

\subsection{Change of Frame}

\begin{figure}
\begin{center}
\includegraphics[width=\linewidth]{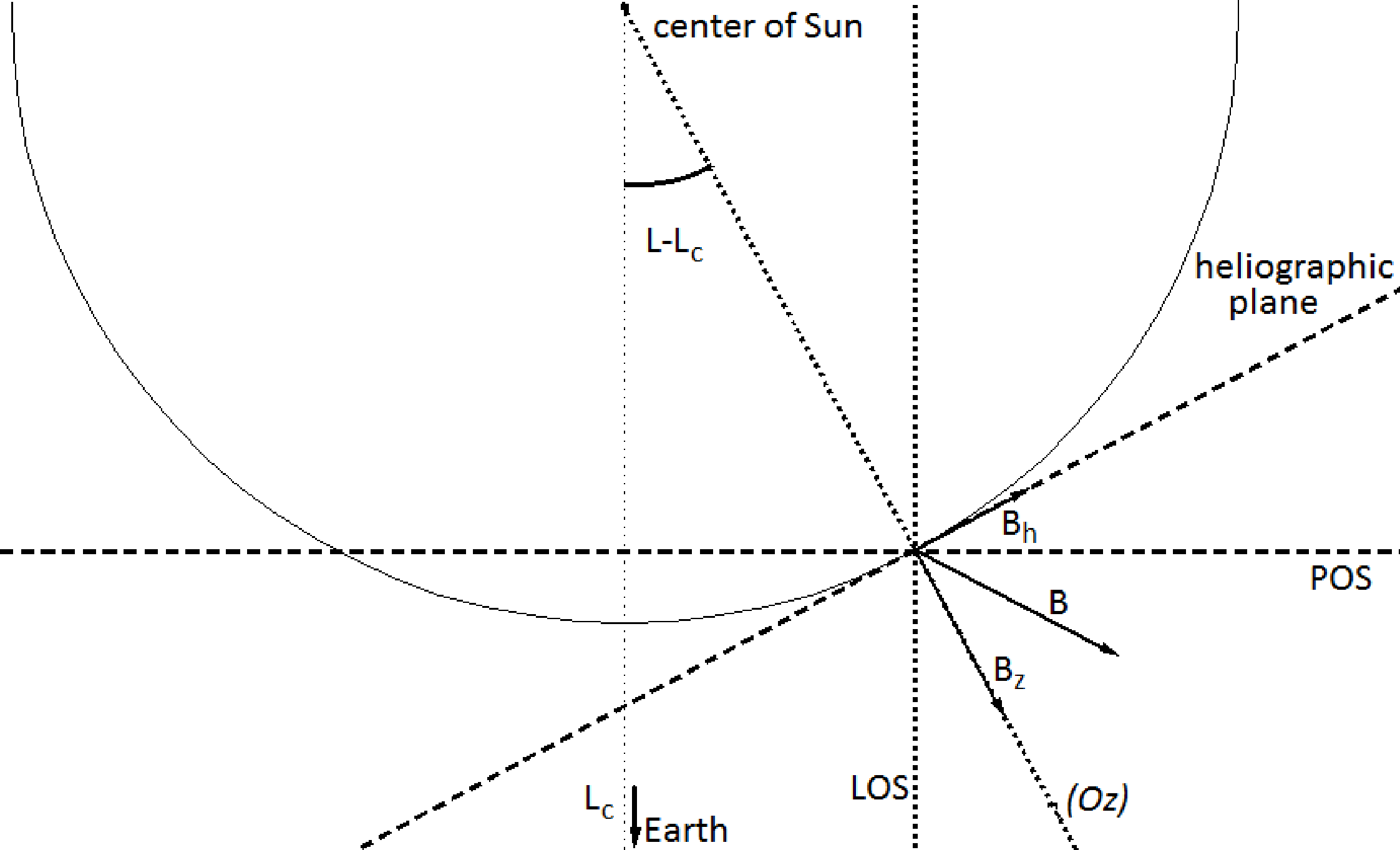}
\end{center}
\caption{The heliographic coordinates: line emission used for spectropolarimetry comes from one layer of the Sun (circle). The magnetic field and electric current density are then calculated in the heliographic coordinates, as the drawing shows: the horizontal component is in the heliographic plane, which is tangent to the considered layer of the Sun, and the vertical component is perpendicular to that, along the (Oz) axis. The X-ray emissions are taken in the plane-of-sky (POS), i.e. in the plane perpendicular to the line-of-sight (LOS).}
\label{geometry}
\end{figure}

To compare the X-ray emissions with magnetic field and current density maps, the frame of reference must be carefully chosen. Indeed, the X-ray images are created in the cartesian coordinates (Plane-Of-Sky), whereas the magnetic maps were taken and derivated in the heliographic system (heliographic plane, see figure \ref{geometry}). We chose to use in the paper the cartesian coordinates, since X-ray sources vary in timescales of the order of seconds, whereas the magnetic field maps are obtained at a cadence of 12 minutes.

The relation between ($x$,$y$), the cartesian coordinates of a point and its heliographic coordinates ($L$,$b$) is given by:
\begin{equation}
\begin{cases}
x = \sin(L-L_{C}) \cos(b) \\
y = \sin(b) \cos(B_{C}) - \cos(L-L_{C}) \cos(b) \sin(B_{C})
\end{cases}
\label{change_of_frame}
\end{equation}
with $L$ the longitude of the point relative to the central meridian, $b$ the latitude of the point, $B_{C}$ and $L_{C}$ the latitude and longitude of the center of the solar disk at the time of observation.

The change of frame also modifies the pixel size. In the case of an active region located near the disk center, several approximations can be made and the pixel size is described by:
\begin{equation}
\begin{cases}
px_{car} \approx px_{helio}\cos(L-L_{C})\\
py_{car} \approx px_{helio} ( \cos(b) \cos(B_{C}) \\
 \  \  \  \ \  \ \ \ \ \ \ \ \ \ \ \ \ \ \ \ \ + \cos(L-L_{C}) \sin(b) \sin(B_{C}) )
\end{cases}
\end{equation}
with ($px_{car}$,$py_{car}$) the pixel size in the cartesian coordinates (in the POS), and ($px_{helio}$,$py_{helio}$) the pixel size in the heliographic coordinates (in the heliographic plane).

\subsection{Time gap between different observations}

The second problem when comparing X-ray emissions and magnetic field maps is the time gap between the different observations. We use the time of the X-ray images as a reference since X-ray sources vary faster than the magnetic field, and also because we generally have more images in X-rays than maps of the magnetic field.

The time gap between the X-ray images and the magnetic maps must be taken into account because the rotation of the solar surface is not negligible. It has been corrected by setting the observation time of the X-ray image when looking for $B_{C}$ and $L_{C}$ in the ephemerids to use the equation \ref{change_of_frame}. Note that we used the get\_sun procedure in the SolarSoft to get the ephemerids ($B_{C}$ and $L_{C}$ at time of observation).

\bibliographystyle{aa} % style aa.bst
\bibliography{zmabiblio} % your references Yourfile.bib

\end{document}